\documentclass[pre,twocolumn,floatfix,superscriptaddress,showpacs,showkeys,preprintnumbers,amsmath,amssymb]{revtex4}

\usepackage{graphicx}
\usepackage{dcolumn}
\usepackage{bm}

\begin{document}

\preprint{APS/123-QED}

\title{Bifurcations and sudden current change in ensembles of\\ 
classically chaotic ratchets} 
\author{Anatole Kenfack}
\affiliation{Max-Planck-Institut f\"ur Physik Komplexer Systeme, 
N\"othnitzer Strasse 38, D-01187 Dresden, Germany}
\author{Sean Sweetnam}
\affiliation{Department of Physics and Astronomy, Carleton College, 
Northfield, Minnesota 55057}
 \author{Arjendu K. Pattanayak}
\affiliation{Department of Physics and Astronomy, Carleton College, 
Northfield, Minnesota 55057}

\date{\today}

\begin{abstract}
In \prl 84, 258 (2000), Mateos conjectured that current reversal 
in a classical deterministic ratchet is associated with bifurcations 
from chaotic to periodic regimes. This is based on the comparison 
of the current and the bifurcation diagram as a function of a given
parameter for a periodic asymmetric potential. Barbi and Salerno, in 
\pre 62, 1988 (2000), have further investigated this claim and argue
that, contrary to Mateos' claim, current reversals can occur also in 
the absence of bifurcations. Barbi and Salerno's studies are based 
on the dynamics of one particle rather than the statistical mechanics of 
an ensemble of particles moving in the chaotic system. The behavior of
ensembles can be quite different, depending upon their characteristics,
which leaves their results open to question. In this paper we present 
results from studies showing how the current depends on the details 
of the ensemble used to generate it, as well as conditions for 
convergent behavior (that is, independent of the details of the 
ensemble). We are then able to present the converged current as a 
function of parameters, in the same system as Mateos as well as
Barbi and Salerno. We show evidence for current reversal without 
bifurcation, as well as bifurcation without current reversal. We 
conjecture that it is appropriate to correlate abrupt changes 
in the current with bifurcation, rather than current reversals, 
and show numerical evidence for our claims.
\end{abstract}


\maketitle

\section{Introduction}
The transport properties of nonlinear non-equilibrium dynamical 
systems are far from well-understood\cite{dorfman}. 
Consider in particular so-called ratchet systems which are asymmetric 
periodic potentials where an ensemble of particles experience 
directed transport\cite{revratchet1,revratchet2}. The origins of the 
interest in this lie in considerations about extracting useful 
work from unbiased noisy fluctuations as seems to happen in biological 
systems\cite{feynman1966,bio-ratchet-review}. Recently attention has
been focused on the behavior of deterministic chaotic
ratchets\cite{chaotic-ratchet1,mateos,barbi,bandyo,chen,son} as well as
Hamiltonian ratchets\cite{roland,flach}. 

Chaotic systems are defined as those which are sensitively dependent 
on initial conditions. Whether chaotic or not, the behavior of nonlinear 
systems -- including the transition from regular to chaotic behavior -- 
is in general sensitively dependent on the parameters of the system. 
That is, the phase-space structure is usually relatively complicated, 
consisting of stability islands embedded in chaotic seas, for examples, 
or of simultaneously co-existing attractors. This can change 
significantly as parameters change. For example, stability islands can 
merge into each other, or break apart, and the chaotic sea itself may 
get pinched off or otherwise changed, or attractors can change 
symmetry or bifurcate. This means that the transport properties 
can change dramatically as well.  A few years ago, 
Mateos\cite{mateos} considered a specific ratchet model with a 
periodically forced underdamped particle. He looked at an ensemble 
of particles, specifically the velocity for the particles, 
averaged over time and the entire ensemble. He showed that this 
quantity, which is an intuitively reasonable definition of `the 
current', could be either positive or negative depending on the 
amplitude $a$ of the periodic forcing for the system.  At the 
same time, there exist ranges in $a$ where the trajectory of an 
individual particle displays chaotic dynamics. Mateos conjectured a 
connection between these two phenomena, specifically that the 
reversal of current direction was correlated with a bifurcation 
from chaotic to periodic behavior in the trajectory dynamics. 
Even though it is unlikely that such a result would be universally 
valid across all chaotic deterministic ratchets, it would still be 
extremely useful to have general heuristic rules such as this. 
These organizing principles would allow some handle on characterizing 
the many different kinds of behavior that are possible in such systems. 

A later investigation\cite{barbi} of the Mateos conjecture by Barbi and 
Salerno, however, argued that it was not a valid rule even in the
specific system considered by Mateos. They presented results showing 
that it was possible to have current reversals in the absence of 
bifurcations from periodic to chaotic behavior. They proposed an
alternative origin for the current reversal, suggesting it was related
to the different stability properties of the rotating periodic orbits 
of the system. These latter results seem fundamentally sensible. However, 
this paper based its arguments about currents on the behavior of a {\em 
single} particle as opposed to an ensemble. This implicitly assumes
that the dynamics of the system are ergodic. This is not true in general 
for chaotic systems of the type being considered. In particular, there 
can be extreme dependence of the result on the statistics of the 
ensemble being considered. This has been pointed out in earlier 
studies~\cite{chaotic-ratchet1} which laid out a detailed methodology 
for understanding transport properties in such a mixed regular and 
chaotic system. Depending on specific parameter value, the particular 
system under consideration has multiple coexisting periodic or 
chaotic attractors or a mixture of both. It is hence 
appropriate to understand how a probability ensemble might behave in
such a system.  The details of the dependence on the ensemble are
particularly relevant to the issue of the possible experimental 
validation of these results, since experiments are always conducted, by
virtue of finite-precision, over finite time and finite ensembles. 
It is therefore interesting to probe the results of Barbi and 
Salerno with regard to the details of the ensemble used, and 
more formally, to see how ergodicity alters our considerations 
about the current, as we do in this paper. 

We report here on studies on the properties of the current in a 
chaotic deterministic ratchet, specifically the same system as 
considered by Mateos\cite{mateos} and Barbi and Salerno\cite{barbi}. 
We consider the impact of different kinds of ensembles of particles 
on the current and show that the current depends significantly on 
the details of the initial ensemble. We also show that it is 
important to discard transients in quantifying the current.  This 
is one of the central messages of this paper: Broad heuristics 
are rare in chaotic systems, and hence it is critical to understand 
the ensemble-dependence in any study of the transport properties of 
chaotic ratchets. Having established this, we then proceed to discuss 
the connection between the bifurcation diagram for individual 
particles and the behavior of the current. We find that while we 
disagree with many of the details of Barbi and Salerno's results, 
the broader conclusion still holds. That is, it is indeed possible 
to have current reversals in the absence of bifurcations from 
chaos to periodic behavior as well as bifurcations without any 
accompanying current reversals. 

The result of our investigation is therefore that the transport
properties of a chaotic ratchet are not as simple as the initial 
conjecture. However, we do find evidence for a generalized version 
of Mateos's conjecture. That is, in general, bifurcations for 
trajectory dynamics as a function of system parameter seem to be 
associated with abrupt changes in the current. Depending on the 
specific value of the current, these abrupt changes may lead the 
net current to reverse direction, but not necessarily so. 

We start below with a preparatory discussion necessary to understand 
the details of the connection between bifurcations and current 
reversal, where we discuss the potential and phase-space for single 
trajectories for this system, where we also define a bifurcation 
diagram for this system. In the next section, we discuss the subtleties 
of establishing a connection between the behavior of individual 
trajectories and of ensembles. After this, we are able to compare 
details of specific trajectory bifurcation curves with current curves, 
and thus justify our broader statements above, after which we conclude. 

\section{Regularity and chaos in single-particle ratchet dynamics}

The goal of these studies is to understand the behavior of general 
chaotic ratchets. The approach taken here is that to discover heuristic
rules we must consider specific systems in great detail before 
generalizing. We choose the same $1$-dimensional ratchet considered 
previously by  Mateos\cite{mateos}, as well as Barbi and 
Salerno\cite{barbi}. We consider an ensemble of particles moving in an 
asymmetric periodic potential, driven by a periodic time-dependent 
external force, where the force has a zero time-average. There is no 
noise in the system, so it is completely deterministic, although 
there is damping.  The equations of motion for an individual trajectory 
for such a system are given in dimensionless variables by 
\begin{equation}
\label{Eq:dyn}
\ddot x + b\dot x +\frac{d V(x)}{d t} = a \cos (\omega t)
\end{equation}
where the periodic asymmetric potential can be written in the form
\begin{equation}
V(x) = C - \frac{1}{4 \pi^2\delta}\bigg [ \sin[2 \pi(x- x_0)] +
\frac{1}{4} \sin [ 4\pi (x -x_0)] \bigg ].
\end{equation}
In this equation $C,x_0$ have been introduced for convenience such that 
one potential minimum exists at the origin with $V(0) =0$ and the term
$\delta = \sin(2\pi|x_0|) + \frac{1}{4} \sin(4 \pi|x_0|)$.

\begin{figure}[htb]
{\includegraphics[width=8.0cm,height=10.cm,clip]
{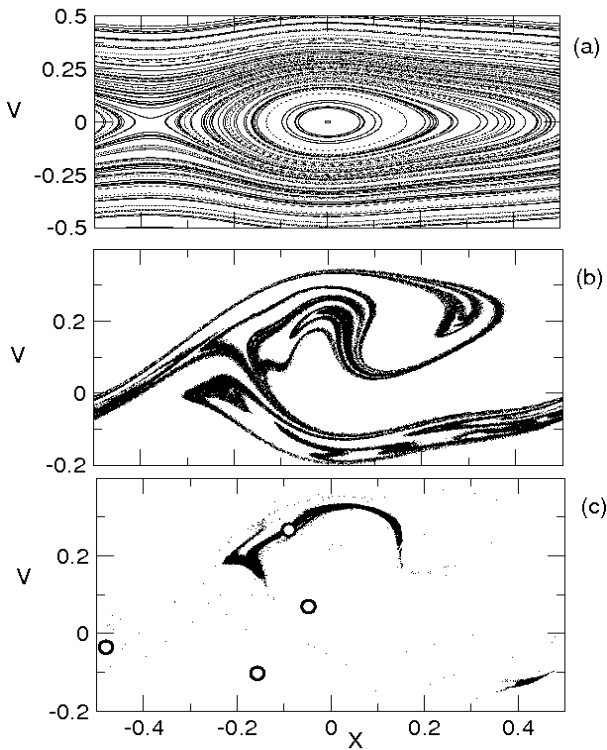}}
\caption{\label{figure1} 
(a) Classical phase space for the unperturbed 
system. For $\omega=0.67$, $b=0.1$, two chaotic attractors emerge with 
$a=0.11$ (b) $a=0.155$ (c) and a period four attractor consisting 
of the four centers of the  circles with $a=0.08125$.}
\end{figure}

The phase-space of the undamped undriven ratchet -- the system
corresponding to the unperturbed potential $V(x)$ -- looks like a series 
of asymmetric pendula. That is, individual trajectories have one of 
following possible time-asymptotic behaviors: (i) Inside the potential 
wells, trajectories and all their properties oscillate, leading to zero 
net transport. Outside the wells, the trajectories either (ii) librate 
to the right or (iii) to the left, with corresponding net transport 
depending upon initial conditions. There are also (iv) trajectories 
on the separatrices between the oscillating and librating orbits, 
moving between unstable fixed points in infinite time, as well as the 
unstable and stable fixed points themselves, all of which constitute a 
set of negligible measure.

When damping is introduced via the $b$-dependent term in Eq.~\ref{Eq:dyn}, 
it makes the stable fixed points the only attractors for the system. 
When the driving is turned on, the phase-space becomes chaotic with the
usual phenomena of intertwining separatrices and resulting homoclinic
tangles. The dynamics of individual trajectories in such a system are now 
very complicated in general and depend sensitively on the choice of 
parameters and initial conditions. We show snapshots of the development
of this kind of chaos in the set of Poincar\'e sections 
Fig.~(\ref{figure1}b,c) together with a period-four orbit represented by 
the center of the circles.

A broad characterization of the dynamics of the problem as a function of 
a parameter ($a,b$ or $\omega$) emerges in a bifurcation diagram. This 
can be constructed in several different and essentially equivalent ways. 
The relatively standard form that we use proceeds as follows: First 
choose the bifurcation parameter (let us say $a$) and correspondingly 
choose fixed values of $b,\omega$, and start with a given value for 
$a =a_{min}$. Now iterate an initial condition, recording the value of 
the particle's position $x(T_P)$ at times $T_p$ from its integrated 
trajectory (sometimes we record $\dot x(T_P)$. This is done 
stroboscopically at discrete times $T_P = n_p*T_\omega$ where 
$T_\omega= \frac{2\pi}{\omega}$ and $n_P$ is an integer $1 \le n_P < M$ 
with $M$ the maximum number of observations made. Of these, discard 
observations at times less than some cut-off time $n_c*T_\omega$ 
and plot the remaining points against $a_{min}$. It must be noted 
that discarding transient behavior is critical to get results which 
are independent of initial condition, and we shall emphasize this 
further below in the context of the net transport or current. 

If the system has a fixed-point attractor then all of the data lie at one 
particular location $x_c$. A periodic orbit with period $j*T_\omega$ 
(that is, with period commensurate with the driving) shows up with 
$M-n_t$ points occupying only $j$ different locations of $x$ 
for $a_{min}$. All other orbits, including periodic orbits of 
incommensurate period result in a simply-connected or multiply-connected 
dense set of points. For the next value $a=a_{min}+da$, the last computed 
value of $x,v$ at $a=a_{min}$ are used as initial conditions, and previously, 
results are stored after cutoff and so on until $a=a_{min}+(j-1)*da=a_{max}$.
That is, the bifurcation diagram is generated by sweeping the 
relevant parameter, in this case $a$, from $a_{min}$ through some 
maximum value $a_{max}$. This procedure is intended to catch all 
coexisting attractors of the system with the specified parameter range. 
Note that several initial conditions are effectively used troughout the 
process, and a bifurcation diagram is not the behavior of a single 
trajectory. We have made several plots, as a test, with different initial 
conditions and the diagrams obtained are identical. We show several 
examples of this kind of bifurcation diagram below, where they are being
compared with the corresponding behavior of the current.

Having broadly understood the wide range of behavior for individual 
trajectories in this system, we now turn in the next section to a 
discussion of the non-equilibrium properties of a statistical ensemble 
of these trajectories, specifically the current for an ensemble.

\section{Ensemble currents}

The current $J$ for an ensemble in the system is defined in an intuitive 
manner by Mateos\cite{mateos} as the time-average of the average velocity 
over an ensemble of initial conditions. That is, an average over several
initial conditions is performed at a given observation time $t_j$ to 
yield the average velocity over the particles 
\begin{equation}
v_j = \frac{1}{N} \sum_{i=1}^N \dot x_i(t_j).
\end{equation}
This average velocity is then further time-averaged; given the 
discrete time $t_j$ for observation this leads to a second sum
\begin{equation}
J = \frac{1}{M} \sum_{j=1}^M v_j
\end{equation}
where $M$ is the number of time-observations made. 

\begin{figure}[htb]
{\includegraphics[width=8.0cm,height=6.cm,clip]
{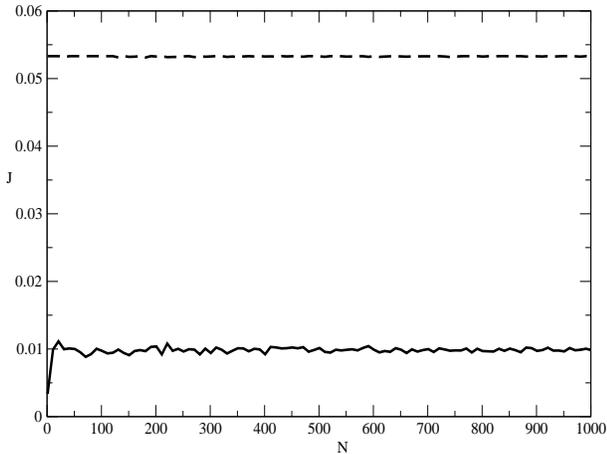}}

\caption{ \label{figure2}Current $J$ versus the number of trajectories $N$ 
for $\omega=0.67$; dashed lines  correspond to a regular motion with 
$a=0.12$  while  solid lines correspond to a chaotic motion with $a=0.08$. 
Note that a single trajectory is sufficient for a regular motion while the 
convergence in the chaotic case is only obtained if the $N$ exceeds a 
certain threshold,  $N\ge N_{thr}=100$.}
\end{figure}

For this to be a relevant quantity to compare with bifurcation diagrams, 
$J$ should be independent of the quantities $N,M$ but still strongly 
dependent on $a,b,\omega$. A further parameter dependence that is being 
suppressed in the definition above is the shape and location of the 
ensemble being used. That is, the transport properties of an ensemble 
in a chaotic system depend in general on the part of the phase-space 
being sampled.  It is therefore important to consider many different 
initial conditions to generate a current.
The first straightforward result we show in Fig.~(\ref{figure2}) is
that in the case of chaotic trajectories, a single trajectory easily 
displays behavior very different from that of many trajectories.
However, it turns out that in the regular regime, it is possible to 
use a single trajectory to get essentially the same result as obtained 
from many trajectories.

\begin{figure}[htb]
{\includegraphics[width=8.0cm,height=6.cm,clip]
{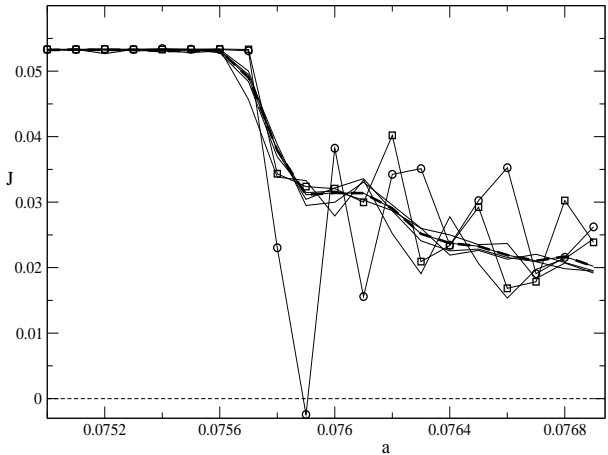}}
\caption{\label{figure3}
Current $J$ versus $a$ for different set of trajectories $N$; $N=1$ 
(circles), $N=10$ (square) and $N=100$ (dashed lines). Note that a
single trajectory suffices in the regular regime where all the curves 
match. In the chaotic regime, as $N$ increases, the curves converge towards 
the dashed one.}
\end{figure}

Further consider the bifurcation diagram in Fig.~(\ref{figure3}) where 
we superimpose the different curves resulting from varying the number 
of points in the initial ensemble. First, the curve is significantly 
smoother as a function of $a$ for larger $N$. Even more relevant is the 
fact that the single trajectory data ($N=1$) may show current reversals 
that do not exist in the large $N$ data. 

Also, note that single-trajectory current values are typically 
significantly greater than ensemble averages. This arises from the fact 
that an arbitrarily chosen ensemble has particles with idiosyncratic 
behaviors which often average out. As our result, with these ensembles 
we see typical $J\approx 0.01$ for example, while Barbi and Salerno report 
currents about $10$ times greater. However, it is not true that only a
few trajectories dominate the dynamics completely, else there would not
be a saturation of the current as a function of $N$. All this is clear 
in Fig.~(\ref{figure3}). We note that the {\bf net} drift of an ensemble 
can be a lot closer to $0$ than the behavior of an individual trajectory.

It should also be clear that there is a dependence of the current on the 
location of the initial ensemble, this being particularly true for 
small $N$, of course. The location is defined by its centroid 
$\langle x\rangle,\langle p\rangle$. For $N=1$, it is trivially true 
that the initial location matters to the asymptotic value of the 
time-averaged velocity, given that this is a non-ergodic and chaotic 
system. Further, considering a Gaussian ensemble, say, the width of 
the ensemble also affects the details of the current, and can show, 
for instance, illusory current reversal, as seen in 
Figs.~(\ref{Current-bifur1},\ref{Current-bifur2}) for example. 
Notice also that in Fig.~(\ref{Current-bifur1}), at $a\approx 0.065$ 
and $a\approx 0.15$, the deviations between the different ensembles is
particularly pronounced. These points are close to bifurcation points
where some sort of symmetry breaking is clearly occuring, which underlines 
our emphasis on the relevance of specifying ensemble 
characteristics in the neighborhood of unstable behavior. However, why 
these specific bifurcations should stand out among all the bifurcations
in the parameter range shown is not entirely clear. 

To understand how to incorporate this knowledge into calculations of the
current, therefore, consider the fact that if we look at the classical 
phase space for the Hamiltonian or underdamped $(b=0)$ motion, we see 
the typical structure of stable islands embedded in a chaotic
sea which have quite complicated behavior\cite{roland}. In such a 
situation, the dynamics always depends on the location of the initial 
conditions. However, we are not in the Hamiltonian situation when the 
damping is turned on -- in this case, the phase-space consists in 
general of attractors. That is, if transient behavior is discarded, 
the current is less likely to depend significantly on the location of 
the initial conditions or on the spread of the initial conditions. 

In particular, in the chaotic regime of a non-Hamiltonian system, the 
initial ensemble needs to be chosen larger than a certain threshold to 
ensure convergence. However, in the regular regime, it is not important 
to take a large ensemble and a single trajectory can suffice, as long 
as we take care to discard the transients. That is to say, in the 
computation of currents, the definition of the current needs to 
be modified to: 
\begin{equation}
J = \frac{1}{M -n_c} \sum_{j=n_c}^M v_j
\end{equation}
where $n_c$ is some empirically obtained cut-off such that we get a
converged current (for instance, in our calculations, we obtained
converged results with $n_c=1000,M=20000$). When this modified form 
is used, the convergence (ensemble-independence) is more rapid as a 
function of $N, M$ and the width of the intial conditions. 

Armed with this background, we are now finally in a position to compare 
bifurcation diagrams with the current, as we do in the next section. 

\section{The relationship between bifurcation diagrams and ensemble currents}

\begin{figure}[htb]
{\includegraphics[width=8cm,clip]
{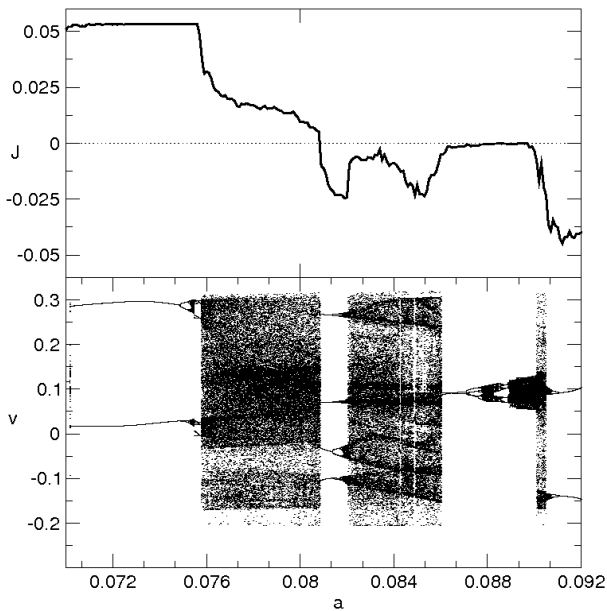}}
\caption{\label{figure5}
For $\omega=0.67$ and $b=0.1$, we plot 
current (upper) with $N=1000$ and bifurcation diagram (lower) 
versus $a$.  Note that there is a {\bf single} current reversal 
while there are many bifurcations visible in the same parameter range.}
\end{figure}

Our results are presented in the set of figures Fig.~(\ref{figure5}) -- 
Fig.~(\ref{rev-nobifur}), in each of which we plot both the ensemble 
current and the bifurcation diagram as a function of the parameter $a$. 
The main point of these numerical results can be distilled into a series 
of heuristic statements which we state below; these are labelled with 
Roman numerals.

Consider Fig.~(\ref{figure5}), which shows the parameter range $a
=\{0.07,0.094\}$ chosen relatively arbitrarily. In this figure, we see
several period-doubling bifurcations leading to order-chaos transitions,
such as for example in the approximate ranges $a=\{0.075,0.076\},
\{0.08,0.082\},\{0.086,0.09\}$. However, there is only one instance of
current-reversal, at $a\approx 0.08$. Note, however, that the current 
is not without structure -- it changes fairly dramatically as a function 
of parameter. 

\begin{figure}[htb]
{\includegraphics[width=8cm,clip]
{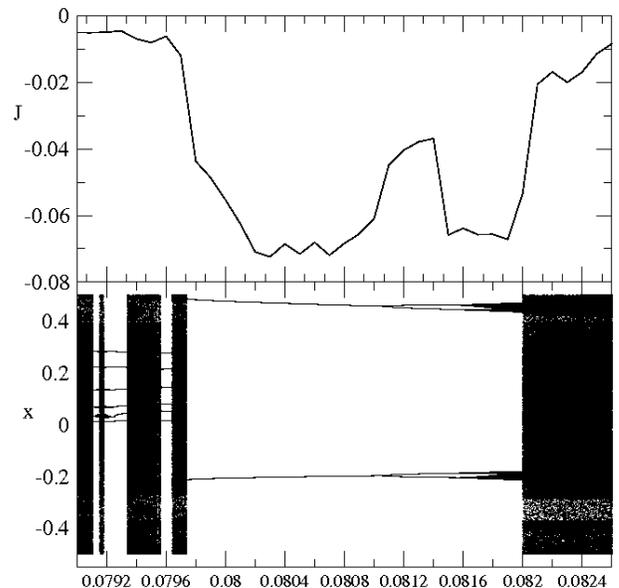}}
\caption{\label{figure6} For $\omega=0.603$ and $b=0.1$, plotted 
are current (upper) and bifurcation diagram (lower) versus $a$ with 
$N=1000$. Notice the current stays consistently below $0$.}
\end{figure}

This point is made even more clearly in Fig.~(\ref{figure6}) where the
current remains consistently below $0$, and hence there are in fact, no
current reversals at all. Note again, however, that the current has
considerable structure, even while remaining negative.
\begin{figure}[htb]
{\includegraphics[width=8cm,clip]
{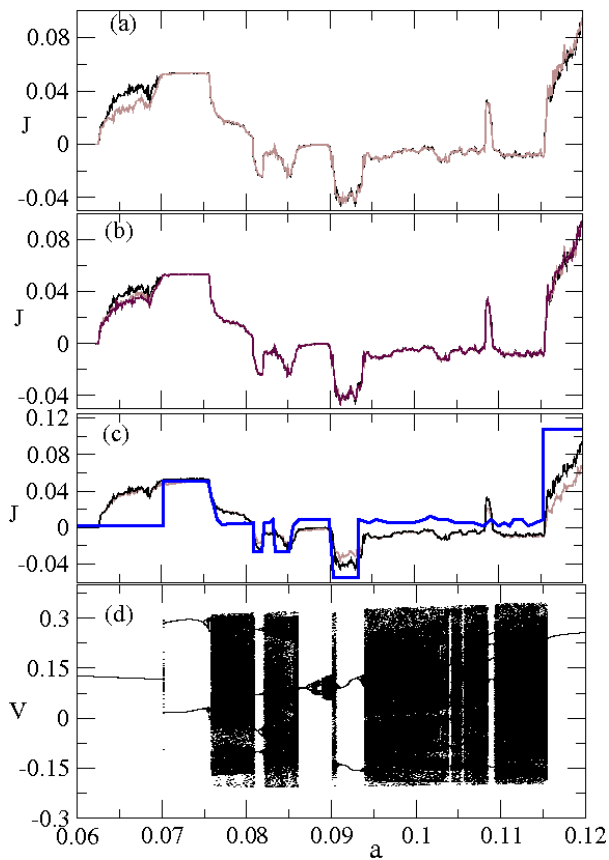}}
\caption{\label{Current-bifur1} Current and bifurcations 
versus $a$. In (a) and (b) we show ensemble dependence, 
specifically in (a) the black curve is for an ensemble of trajectories
starting centered at the stable fixed point $(0,0)$ with a
root-mean-square Gaussian width of $0.25$, and the brown curve for
trajectories starting from the unstable fixed point $(-0.375,0)$ and of
width $0.25$. In (b), all ensembles are centered at the stable fixed
point, the black line for an ensemble of width $0.25$, brown a width of
$0.5$ and maroon with width $1.0$.
(c) is the comparison of the current $J$ without transients (black) 
and with transients (brown) along with the single-trajectory results
in blue (after Barbi and Salerno). The initial conditions for the ensembles
are centered at $(0,0)$ with a mean root square gaussian of width 
$0.25$.  (d) is the corresponding bifurcation diagram.}
\end{figure}
It is possible to find several examples of this at different parameters, 
leading to the negative conclusion, therefore, that {\bf (i) not all 
bifurcations lead to current reversal}.
However, we are searching for positive correlations, and at this point 
we have not precluded the more restricted statement that all current 
reversals are associated with bifurcations, which is in fact Mateos' 
conjecture.

We therefore now move onto comparing our results against the specific 
details of Barbi and Salerno's treatment of this conjecture. 
In particular, we look at their Figs.~(2,3a,3b), where they scan the 
parameter region $b=0.1,\omega=0,67,a=\in (0.0,0.24)$. The distinction 
between their results and ours is that we are using {\em ensembles} of 
particles, and are investigating the convergence of these results as a 
function of number of particles $N$, the width of the ensemble in 
phase-space, as well as transience parameters $n_c,M$. 

\begin{figure}[htbp]
{\includegraphics[width=8cm,clip]
{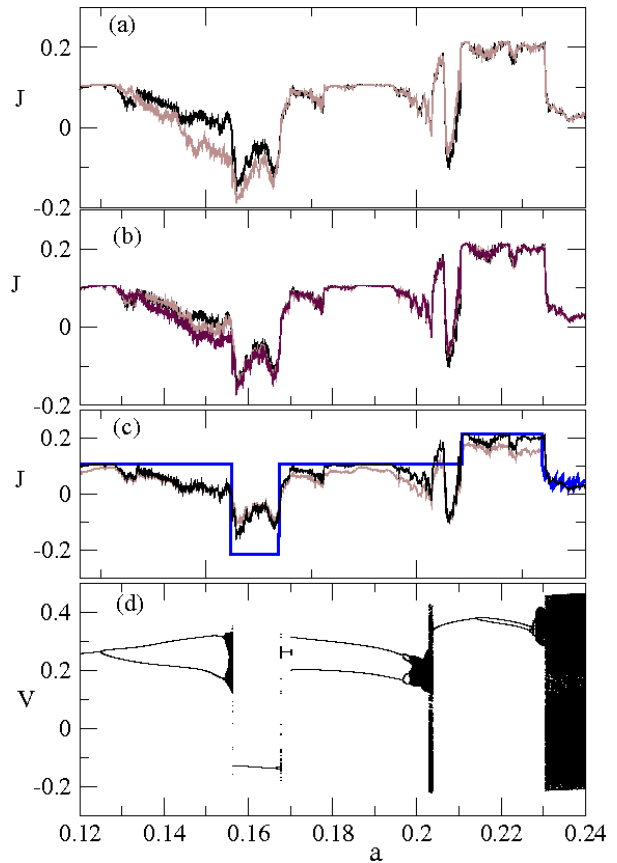}}
\caption{\label{Current-bifur2}
Same as Fig.~(\ref{Current-bifur1}) except for the range of $a$ considered.}
\end{figure}

Our data with larger $N$ yields different results in general, as we 
show in the recomputed versions of these figures, presented here in 
Figs.~(\ref{Current-bifur1},\ref{Current-bifur2}). Specifically,
(a) the single-trajectory results are, not surprisingly, cleaner
and can be more easily interpreted as part of transitions in the 
behavior of the stability properties of the periodic orbits. 
The ensemble results on the other hand, even when converged, show 
statistical roughness. (b) The ensemble results are consistent with 
Barbi and Salerno in general, although disagreeing in several details. 
For instance, (c) the bifurcation at $(a\approx 0.07)$ has a much 
gentler impact on the ensemble current, which has been growing for 
a while, while the single-trajectory result changes abruptly. 
Note, (d) the very interesting fact that the single-trajectory current 
completely misses the bifurcation-associated spike at $(a\approx 0.11)$.
Further, (e) the Barbi and Salerno discussion of the behavior of the 
current in the range $a \in (0.14,0.18)$ is seen to be flawed -- our 
results are consistent with theirs, however, the current changes are 
seen to be consistent with bifurcations despite their statements to the
contrary. On the other hand (f), the ensemble current shows a case 
[in Fig.~(\ref{Current-bifur2}), at $a>0.2$] of current reversal that 
does not seem to be associated with bifurcations. 
In this spike, the current abruptly drops below $0$ and then rises 
above it again. The single trajectory current completely ignores this 
particular effect, as can be seen. The bifurcation diagram indicates 
that in this case the important transitions happen either before or 
after the spike. 

All of this adds up to two statements: The first is a reiteration of the
fact that there is significant information in the ensemble current that 
cannot be obtained from the single-trajectory current. The second is
that the heuristic that arises from this is again a negative conclusion, 
that {\bf (ii) not all current reversals are associated with bifurcations.}
Where does this leave us in the search for `positive' results, that is,
useful heuristics? One possible way of retaining the Mateos conjecture 
is to weaken it, i.e. make it into the statement that {\bf (iii) {\em most} 
current reversals are associated with bifurcations.}

\begin{figure}[htb]
{\includegraphics[width=8cm,clip]
{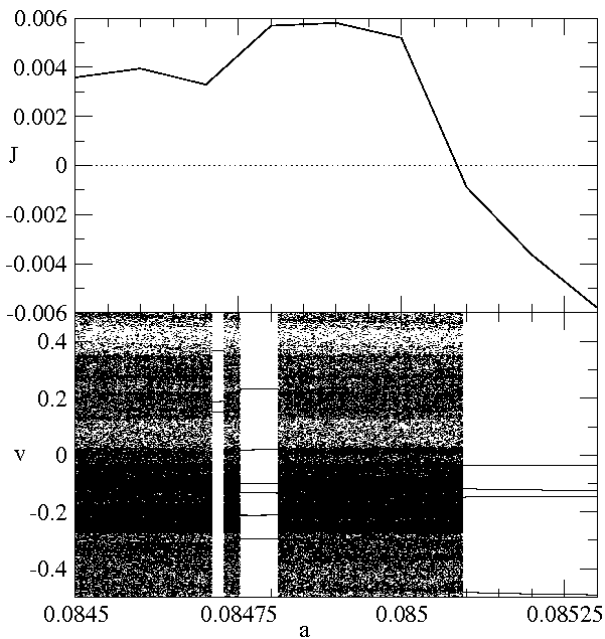}}
\caption{\label{hidden-bifur} 
For $\omega=0.6164$ and $b=0.1$, plotted are current (upper) and 
bifurcation diagram (lower) versus $a$ with $N=1000$. Note in 
particular in this figure that eyeball tests can be misleading. 
We see reversals without bifurcations in (a) whereas the zoomed
version (c) shows that there are windows of periodic and chaotic 
regimes. This is further evidence that jumps in the current 
correspond in general to bifurcation.}
\end{figure}
\begin{figure}[htb]
{\includegraphics[width=8cm,clip]
{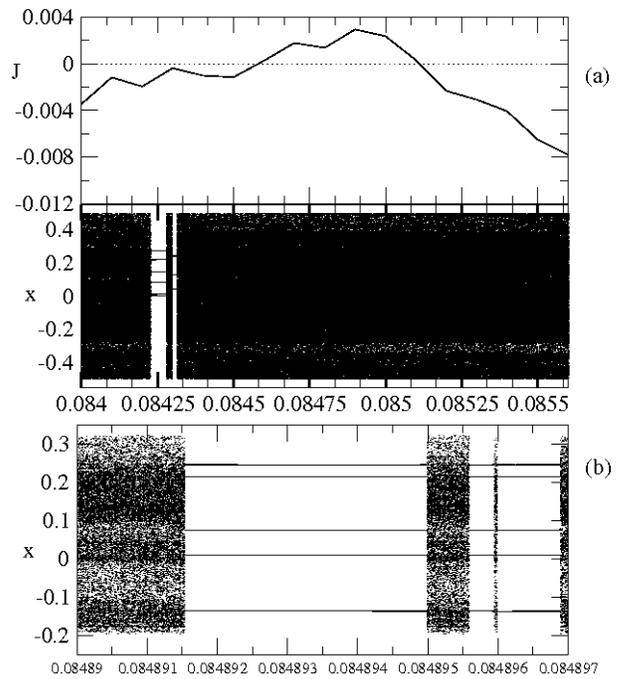}}
\caption{\label{rev-nobifur} For $\omega=0.67$ and $b=0.11$, 
current (upper) and bifurcation diagram (lower) versus $a$.}
\end{figure}

However, a {\bf different} rule of thumb, previously not proposed, emerges 
from our studies. This generalizes Mateos' conjecture to say that 
{\bf (iv) bifurcations correspond to sudden current changes (spikes or 
jumps)}. Note that this means these changes in current are not necessarily 
reversals of direction. 
If this current jump or spike goes through zero, this coincides with a 
current reversal, making the Mateos conjecture a special case. 
The physical basis of this argument is the fact that ensembles of 
particles in chaotic systems {\em can} have net directed transport 
but the details of this behavior depends relatively sensitively on 
the system parameters. This parameter dependence is greatly exaggerated 
at the bifurcation point, when the dynamics of the underlying 
single-particle system undergoes a transition -- a period-doubling 
transition, for example, or one from chaos to regular behavior.
Scanning the relevant figures, we see that this is a very useful rule 
of thumb. For example, it completely captures the behaviour of 
Fig.~(\ref{figure6}) which cannot be understood as either an example 
of the Mateos conjecture, or even a failure thereof. As such, this 
rule significantly enhances our ability to characterize changes 
in the behavior of the current as a function of parameter. 

A further example of where this modified conjecture helps us is in
looking at a seeming negation of the Mateos conjecture, that is, an 
example where we seem to see current-reversal without bifurcation, 
visible in Fig.~(\ref{hidden-bifur}).
The current-reversals in that scan of parameter space seem 
to happen inside the chaotic regime and seemingly independent of
bifurcation.  However, this turns out to be a `hidden' bifurcation 
-- when we zoom in on the chaotic regime, we see hidden periodic
windows. This is therefore consistent with our statement that 
sudden current changes are associated with bifurcations. Each of the 
transitions from periodic behavior to chaos and back provides 
opportunities for the current to spike. 

However, in not all such cases can these hidden bifurcations be 
found.  We can see an example of this in Fig.~(\ref{rev-nobifur}). 
The current is seen to move smoothly across $J = 0$ with seemingly 
no corresponding bifurcations, even when we do a careful zoom on 
the data, as in Fig.~(\ref{hidden-bifur}). However, arguably, 
although subjective, this change is 'close' to the bifurcation 
point. This result, that there are situations where the heuristics 
simply do not seem to apply, are part of the open questions 
associated with this problem, of course. We note, however, that we 
have seen that these broad arguments hold when we vary other 
parameters as well (figures not shown here).

In conclusion, in this paper we have taken the approach that it is 
useful to find general rules of thumb (even if not universally valid) to
understand the complicated behavior of non-equilibrium nonlinear
statistical mechanical systems. In the case of chaotic deterministic
ratchets, we have shown that it is important to factor out
issues of size, location, spread, and transience in computing the
`current' due to an ensemble before we search for such rules, and that
the dependence on ensemble characteristics is most critical near certain
bifurcation points. We have then argued that the following heuristic 
characteristics hold: Bifurcations in single-trajectory behavior often
corresponds to sudden spikes or jumps in the current for an ensemble in
the same system. Current reversals are a special case of this. However,
not all spikes or jumps correspond to a bifurcation, nor vice versa. The
open question is clearly to figure out if the reason for when these
rules are violated or are valid can be made more concrete.

\section{Acknowledgements}
A.K. gratefully acknowledges T. Barsch and Kamal P. Singh for stimulating 
discussions, the Reimar L\"ust grant and financial support from 
the Alexander von Humboldt foundation in Bonn. A.K.P. is grateful to 
Carleton College for the `SIT, Wallin, and Class of 1949' sabbatical 
Fellowships, and to the MPIPKS for hosting him for a sabbatical visit, 
which led to this collaboration. Useful discussions with J.-M. Rost on 
preliminary results are also acknowledged.


\begin{thebibliography}{99}

\bibitem{dorfman} See for example J.~R.~Dorfman, {\em An introduction
to chaos in nonequilibrium statistical mechanics}, Cambridge University
Press, New York (1999).

\bibitem{revratchet1} P. H\"anggi and Bartussek, in Nonlinear physics of 
complex systems, Lecture notes in Physics Vol. 476, edited by J. Parisi, 
S.C. Mueller, and W. Zimmermann (Springer Verlag, Berlin, 1996), 
pp.294-308;
\bibitem{revratchet2} R.D. Asturmian, Science {\bf 276}, 917 (1997); F. 
J\"ulicher, A.~Ajdari, and J.~Prost, Rev.~Mod.~Phys.~{\bf 69}, 1269 
(1997); C. D\"oring, Nuovo Cimento~D{\bf 17}, 685 (1995)

\bibitem{feynman1966}
R.~.P.~Feynmann, R.~B.~Leighton, and M. Sands, The Feynmann Lectures on 
Physics (Addison- Wesley, Reading, MA, 1966), Vol.1, Chap.46

\bibitem{bio-ratchet-review}
R. D. Asturmian, and I. Derenyi, Eur. Biophys. J.~{\bf 27}, 474 (1998). 

\bibitem{chaotic-ratchet1} 
P. Jung, J.~G.~Kissner, and P. H\"anggi, Phys.~Rev.~Lett.~{\bf 76}, 3436 
(1995); N.~Thomas and R.~A.~Thornhill, J.~Phys.~D.~{\bf 31}, 253 (1998).

\bibitem{mateos} Jos{\'e} L. Mateos, \prl {\bf 84}, 258 (2000). 

\bibitem{barbi}Maria Barbi and Mario Salerno, \pre {\bf 62} 1988 (2000).

\bibitem{bandyo}M.~Bandyopadhyay, S.~Dattagupta, and M.~Sanyal, \pre
{\bf 73}, 051108 (2006) 

\bibitem{chen}H.~Chen, Q.~Wang, and Z.~Zheng, \pre {\bf 71}, 031102 (2005)

\bibitem{son} W.-S. Son, I. Kim, Y.-J. Park, and C.-M. Kim, \pre
{\bf 68}, 067201 (2003).

\bibitem{roland}
H. Schanz, M.-F. Otto, R. Ketzmerick, and T. Dittrich,
Phys.~Rev.Lett.~{\bf 87}, 070601 (2001); H. Schanz, T. Dittrich, and R.
Ketzmerick, Phys.~Rev.~E~{\bf 71}, 026228 (2005).

\bibitem{flach}S. Flach, O. Yevtushenko, and Y. Zolotaryuk, 
Phys.~Rev.~Lett.~{\bf 84}, 2358 (2000); O. Yevtushenko, S. Flach, Y.
Zolotaryuk, and A.~A. Ovchinnikov, Europhys.~Lett.~{\bf 54}, 141 (2001);
S.  Denisov et al., Phys.~Rev.~E~{\bf 66}, 041104 (2002)




\end{thebibliography}
\end{document}